\documentclass[12pt]{book}

\usepackage[text={15.5cm,23cm},centering]{geometry}

\usepackage{mathptmx}
\usepackage{graphicx}
\usepackage{chapterbib} 
\usepackage{subfigure}
\usepackage{color}
\usepackage[colorlinks=true, linkcolor=blue]{hyperref}

\usepackage[english]{babel}  

  \author{Cathryn M. Trott (ICRAR-Curtin), Jonathan Pober (Brown University)}

\begin{document}

\mainmatter
\setcounter{chapter}{6}

\chapter{Status of 21~cm interferometric experiments}

\begin{bf}
  \Large{Cathryn M. Trott (ICRAR-Curtin), Jonathan Pober (Brown University)}
\vspace{0.5cm}
\normalsize

Abstract

Interferometric experiments of the reionization era offer the advantages of measuring power in spatial modes with increased sensitivity afforded by multiple independent sky measurements. Here we review early work to measure this signal, current experiments, and future opportunities, highlighting the lessons learned along the way that have shaped the research field and experimental design. In particular, this chapter discusses the history, progress, challenges and forecasts for detection and exploration of the spatial structure of the 21~cm brightness temperature signal in the Epoch of Reionisation using interferometric experiments. We discuss GMRT, PAPER, LOFAR, MWA, and the future HERA and SKA.
\end{bf}

\section{Introduction}

Because they provide both rapid mapping speed and good angular resolution, interferometers have become the preferred instrument for experiments looking to measure the expected spatial fluctuations in the 21~cm signal.
The current instruments hosting such experiments include the Murchison Widefield Array, MWA{\footnote[1]{http://www.mwatelescope.org}} \cite{bowman13_mwascience,tingay13_mwasystem,jacobs16}; the Precision Array for Probing the Epoch of Reionization, PAPER{\footnote[2]{http://eor.berkeley.edu}} \cite{parsons10}; the LOw Frequency ARray, LOFAR{\footnote[3]{http://www.lofar.org}} \cite{vanhaarlem13,patil16}; and the Long Wavelength Array, LWA{\footnote[4]{http://lwa.unm.edu}} \cite{ellingson09}. In the future, we expect the Hydrogen Epoch Reionization Array, HERA \cite{deboer16} and the Square Kilometre Array, SKA-Low \cite{koopmans15}.
Sensitivity predictions for most of the current experiments (e.g. \cite{beardsley13,pober14}) find that they will not be capable of achieving the necessary signal-to-noise to image the 21~cm signal directly (although see \cite{2012MNRAS.425.2964Z} for a study with LOFAR).  As such, most of these experiments are targeting a detection of the 21~cm power spectrum, which can be constrained with higher signal-to-noise compared with an image because the isotropy and homogeneity of the Universe allows the 3D $k$-space power spectrum to be averaged over spherical shells of constant $|k|$.
Even using the power spectrum, typical predictions for the requisite observing time are of order 1,000 hours (see Figure \ref{fig:current}).

However, beyond the need to achieve the requisite sensitivity, experiments are faced with the daunting task of isolating the 21~cm signal from foregrounds that can be up to five orders of magnitude brighter.  While the two can, in principle, be separated by their distinct spectral behavior, the inherently frequency-dependent response of radio interferometers complicates the picture significantly.  In this chapter, we review the challenges faced by current interferometric 21~cm experiments as well as the progress to-date in overcoming them.  The detailed structure of this chapter is as follows.  In \S\ref{sec:early_work}, we present the history of experiments and techniques that led to the design of current 21~cm experiments.  In \S\ref{sec:methodologies}, we discuss the distinct approaches each experiment has developed to overcome the challenges associated with these observations, and in \S\ref{sec:results} we review the current published upper limits on the 21~cm signal strength from these experiments.  In \S\ref{sec:challenges}, we highlight the currently unsolved problems at the forefront of experimental 21~cm cosmology and conclude in \S\ref{sec:prospects} with a discussion of the potential for both current and future experiments to overcome them.

\section{Early work}
\label{sec:early_work}
The origins of the approaches that current experiments are taking to detect the Epoch of Reionization power spectrum can be traced to the development of radio interferometry observational techniques and Cosmic Microwave Background (CMB) analysis methodology.

Radio interferometers measure the cross-correlation of voltages detected with two antennas, extracting the sky signal in a complex-valued dataset that encodes sky emission location and intensity, and as a function of antenna separation vector and frequency \cite{tms}. For small field-of-view instruments (large antenna aperture), the measured signal is well-approximated as the 2D Fourier Transform of the sky brightness, attenuated by the antenna response function (the primary beam).

Motivated by analysis of CMB datasets in the 1990s and 2000s, and the curved nature of full-sky imaging, early discussion of power spectrum estimators used spherical harmonic basis functions to describe the signal and extract optimal estimators \cite{tegmark97}. CMB studies suffer from some of the challenges faced also by EoR experiments: wide fields-of-view, low sensitivity, limited angular resolution, and foreground contamination. Unlike EoR, which is an evolving signal in redshift space, CMB studies are single frequency experiments focussed on angular statistics. As such, the foreground mitigation and treatment approaches of CMB studies are of limited use for EoR studies, which attempt to separate foregrounds from the 21~cm signal using the frequency axis. Nonetheless, the fundamental need to extract a weak signal from complex and highly-contaminated data is shared between the two fields, and Tegmark \cite{tegmark97} used this experience to apply CMB analysis techniques to early EoR methodology development.
Since an interferometer natively measures in Fourier space, there was a transition from the natural basis of curved sky functions (spherical harmonics) to the interferometer measurement space (Fourier modes) in discussion of optimal estimators for EoR science \cite{liu11}.

This work was supported by groundwork laid out for doing EoR power spectra with radio interferometers, including cosmological and unit conversions \cite{morales04,parsons10} and noise considerations for astrophysical parameter estimation with specific future experiments \cite{mcquinn06}. McQuinn and colleagues \cite{mcquinn06} discussed a simple foreground model where fitting of a smooth spectral function could remove their effect cleanly, focussing on array sensitivity as the limiting factor for future experiments.
However, the lack of any real-world experiments attempting the detection meant they failed to realise the extent of foreground spectral contamination.

More sophisticated approaches to foreground modelling and mitigation appeared in the mid-2000s, with \cite{2009ApJ...695..183B} beginning a set of papers that explored the signature of smooth spectrum sources in the EoR power spectrum parameter space. Initially, low-order polynomials were explored to fit and remove these sources. However, lacking a physical motivation for this functional form to robustly separate foregrounds from cosmological signal, polynomials  were replaced with more realistic functions. Ultimately, the likelihood of removing not only foregrounds but also cosmological signal when fitting and subtracting models, particularly considering the large difference in magnitude of the two signals, has steered the research field away from this approach to foreground mitigation.

As part of this better appreciation for the impact of foregrounds, particularly with the knowledge that they are used also for data calibration, \cite{datta10} explored the required accuracy of source models such that foregrounds may be subtracted to a level sufficient to detect the EoR. This work was the first to show the characteristic wedge in power spectrum parameter space, a triangular region in angular and line-of-sight wavenumber space representing the signature of smooth-spectrum sources observed with an interferometer.

\section{Experimental methodologies and current experiments}
\label{sec:methodologies}
\begin{figure}[ht]
\centering
\includegraphics[width=1.0\textwidth]{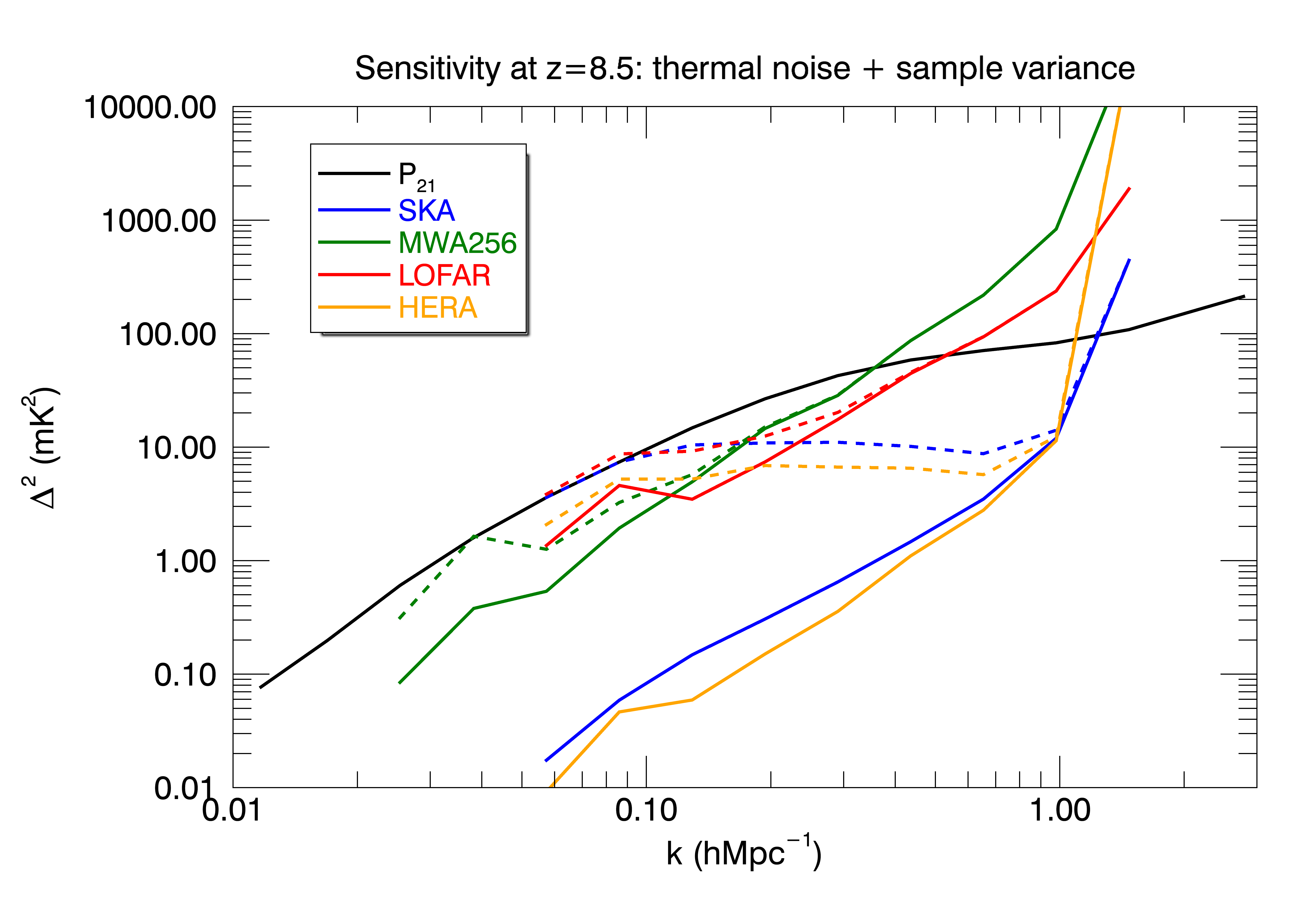}
\caption{Estimated total uncertainty on the dimensionless power spectrum for a 1000 hour observation with several experiments at $z$=8.5 compared with a model input power spectrum (21cmFAST,\cite{mesinger11}), and assuming that modes within the horizon are inaccessible. Solid lines are only thermal noise uncertainty, while dashed lines include sample variance. (Black) Model cosmological signal; (blue) SKA; (green) MWA256; (red) LOFAR; (orange) HERA.}\label{fig:current}
\end{figure}

In this section we introduce the different instruments that have previously taken, or currently are taking and analysing, data for an interferometric EoR experiment. We start by presenting the relevant parameters of the telescopes that these experiments use, highlighting and motivating the different observational and analysis approaches taken by each. Table \ref{table:parameters} lists the location (including latitude), frequency (redshift) range, number of stations/antennas, station diameters, maximum baseline, and field-of-view at 150\, MHz for the relevant instruments. Italicised telescopes are discussed in this Chapter. We also plot the full uncertainties (including sample variance) for a 1000~hour observation at $z$=8.5 (10~MHz bandwidth) for each experiment as a function of spatial wavenumber in Figure \ref{fig:current}. We uniformly assume that the modes within the horizon are inaccessible due to foreground contamination, and note that this is a broad assumption that is not applicable to all experiments (see Chapter 5 for a discussion).
Note also that MWA's and HERA's large fields-of-view gives them access to smaller wavenumbers. This figure also includes a nominal signal strength (black, 21cmFAST, \cite{mesinger11}), but this level is highly uncertain because it depends on the unknown properties of high-z galaxies and the IGM (see Chapter 2). The proximity of the curves to this line highlights the difficulty with predicting the real sensitivity of experiments, particularly in light of the large number of observing hours required to reach an expected detection.
\begin{table}[ht]
\centering
\begin{tabular}{|c||c|c|c|c|c|}
\hline
Facility & Location (Latitude) & Freq. [MHz] ($z$) & $N_{\rm ant}$ & Max. baseline & FOV$_{150}$ \\
\hline
\textit{GMRT} & India (19.1$^{o}$N) & 150--300 (3.7--8.5) & 30 & 30~km & 2.5$^{o}$\\
\textit{MWA} & Australia (26.5$^{o}$S) & 70--90, 135--195 (15--19, 6--10) & 128 & 5~km & 25$^{o}$\\
\textit{LOFAR} & Netherlands (52.9$^{o}$N) & 30--80, 120--190 (17--46, 6--11) & 50--60 & 50~km & 5$^{o}$\\
\textit{PAPER} & South Africa (30.6$^{o}$S) & 110--180 (7--12) & 32--64 & 210~m & 60$^{o}$\\
LEDA\footnote[1]{} & USA (34$^{o}$N) & 45--88 (15--30) & 256+ & $<$10~km & 70$^{o}$\\
21CMA & China (42$^{o}$N) & 50--200 (6--27) & 81 & 6~km & 10$^{o}$\\
\hline \hline
\end{tabular}
\label{table:parameters}
\caption{General parameters for the telescopes undertaking interferometric observations of the Cosmic Dawn and EoR. Italicised telescopes are discussed in this Chapter. $^1$LEDA is a total power experiment using interferometry for data calibration.}
\end{table}
The parameters shown in the table, and the curves shown in Figure \ref{fig:current} motivate and frame the discussion of different experiments in the following sections. Experiments are forced to undertake different approaches to observations and data analysis, because the physical limitations of the systems promote different systematic errors into the forefront for each experiment. There is no silver bullet telescope for undertaking this experiment, however, and after reviewing the main experiments, we discuss the pros and cons of different features.

\subsection{Giant Metrewave Radio Telescope - GMRT}
\label{sec:gmrt}
\begin{figure}[ht]
\centering
\includegraphics[width=0.85\textwidth]{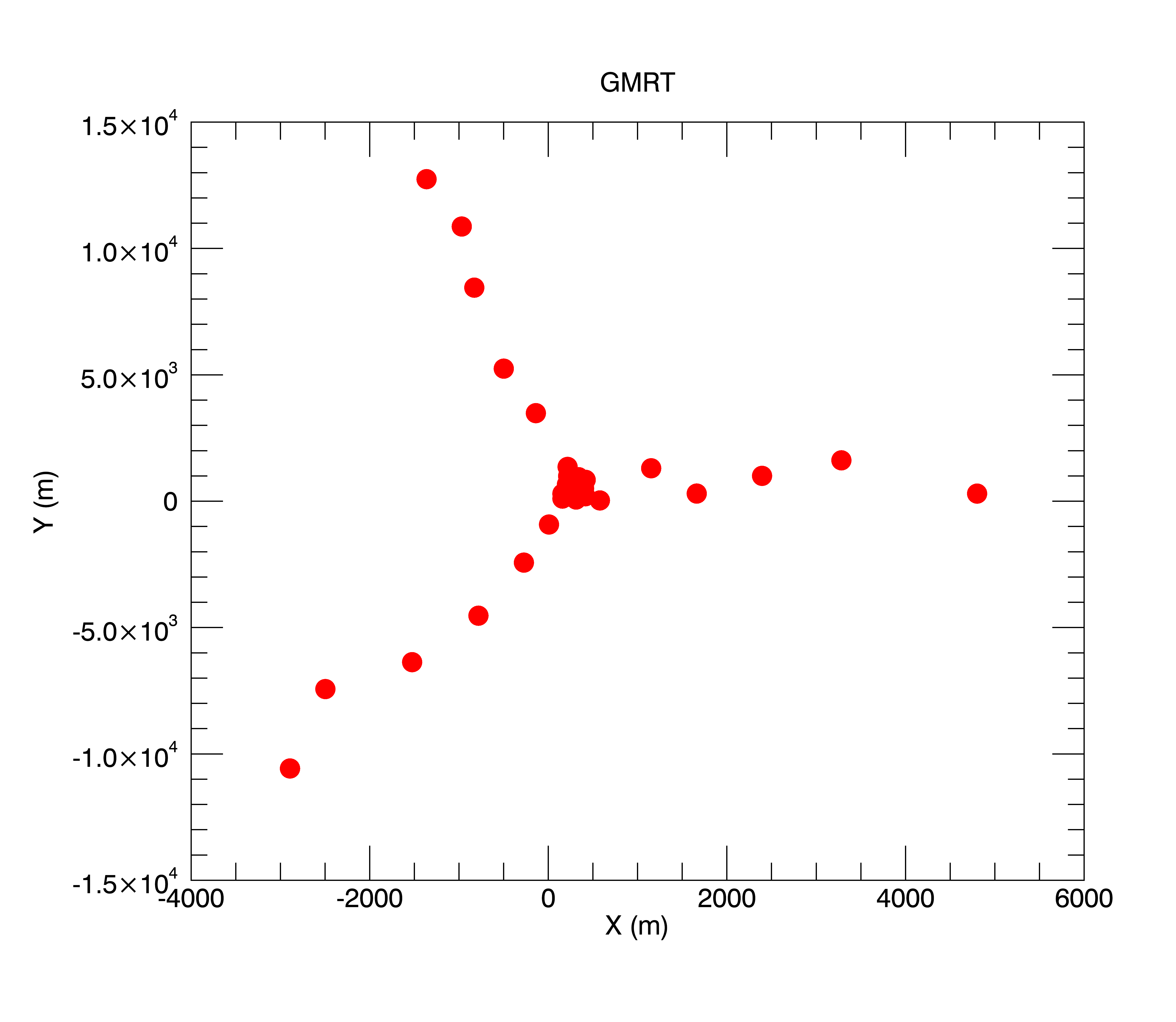}
\caption{Array configuration for the GMRT: thirty 45~m dishes spread over 30~km.}\label{fig:gmrt}
\end{figure}

The GMRT \cite{swarup91} is a Y-shaped array of 30 45~m dishes spread over 25~km in western India. Operating between 50~MHz and 1420~MHz, its 153~MHz receiver has been most used for Reionization studies. Motivated by early work in the post-reionization era (325~MHz and 610~MHz receivers) to statistically detect 21~cm fluctuations and understand the foreground contamination of these data, the low frequency receiver opens the door to exploring the Reionization era. The methodology developed has focused on angular power spectra, measured at a range of frequencies, and pioneered much of the early work to use spectral correlation of foregrounds as a way of treating them. With a lack of short baselines and poor instantaneous $uv$-coverage (Figure \ref{fig:gmrt}), the array is suited to building high resolution foreground models, and computing the foreground angular correlation function (\cite{rana19}).

GMRT work has largely utilised the visibility correlation function, which cross-correlates visibilities to study the spectral and spatial structure of the sky. Visibility correlation functions were also explored for 21CMA analysis \cite{zheng12}. Unlike other experiments, which have cross-correlated interleaved time samples to remove noise power bias, GMRT has usually opted for cross-correlating visibilities from adjacent frequency channels. This has different systematics, with finer spectral resolution required to minimise visibility decorrelation.  However, as is standard practise, time integration is used for reducing noise uncertainty.

During the 2000s, there was a series of papers developing a formalism for use of this visibility correlation function to measure angular modes.
\cite{bharadwaj01} introduced a cross-visibility angular correlation function to measure HI fluctuations post-reionization.
\cite{bharadwaj05} then related the cross-visibility correlation function across baselines and frequencies to the power spectrum of brightness temperature fluctuations, presenting the full formalism and expected results in different epochs. They suggest that the cosmological signal is uncorrelated for frequency channel differences larger than 1~MHz, allowing signal to be `easily distinguished from the continuum sources of contamination'.
\cite{ali08} then extended this formalism to model the expected foreground continuum signatures in the cross-correlation visibility space, and compared with GMRT observations. Their results were hindered by calibration errors, which caused decorrelation of the signal over frequency, but presented the first application of this technique to data.
\cite{datta07} provided an extension of the visibility cross-correlation approach to estimating power spectra to a multi-frequency angular power spectrum (MAPS), utilising decorrelation of signals over frequency to extract information about bubble sizes and distributions as a function of redshift while suppressing the effects of foreground contamination.

\cite{ghosh11} published the first measurement of post-reionization neutral hydrogen fluctuations with GMRT (HI intensity mapping) at $z$=1.32 (610 MHz) and using the MAPS formalism.
They used a fourth-order polynomial to remove smooth foregrounds, in line with early attempts with many experiments to fit a parametric function without physical motivation.
\cite{ghosh11a} then demonstrated improved foreground removal for 610 MHz observations by tapering the primary beam function and reducing sidelobes;
\cite{ghosh12} further extended the work to the reionization epoch using 150 MHz observations to characterize the foregrounds with the MAPS formalism.

Using an alternative analysis to the MAPS formalism, \cite{paciga11} analysed 50h of data at $z$=8.6 with a simple piecewise linear foreground subtraction method and cross-correlation of foreground-subtracted images.  The result was a reported upper limit on the 21~cm signal strength of (70 mK)$^2$.
However, \cite{paciga13} re-analysed the data with a more sophisticated foreground subtraction technique, including a calculation of signal loss due to foreground fitting. The result was an increase in the upper limit to (248 mK)$^2$, indicative of the degree to which signal loss can affect results.

More recently, \cite{chouduri14} published a series of papers introducing and exploring the use of two new optimised power spectrum estimators using visibility correlations: the Tapered Gridded Estimator (TGE) and Bare Estimator (BE). The key concept for the TGE, which has been further discussed in the literature in subsequent papers \cite{2016MNRAS.463.4093C}, is to use a Fourier beam gridding kernel that is larger than the physical beam kernel, thereby decorrelating sources at the edge of the field-of-view.
Note that this approach is not a silver bullet to removing the effect of horizon sources, because their sidelobes remain in the data even if they have been attenuated. Originally developed as angular power spectra as a function of frequency, the TGE work has recently been expanded to use the line-of-sight spatial information \cite{bharadwaj19}. The BE directly squares adjacent visibilities to provide individual measurements of the power, but this has not been used further, possibly due to the large number of visibilities that are accumulated and stored.

Additionally to power spectra, \cite{ali06} predicted the amplitude of a bispectrum signal with GMRT using its shortest baselines by modelling non-linear clustering. They predicted the signal strength to be comparable to the power spectrum and detectable in 100 hours but this project has not been explored observationally with this instrument.

\subsection{Murchison Widefield Array - MWA}
\begin{figure}[ht]
\centering
\includegraphics[width=0.42\textwidth]{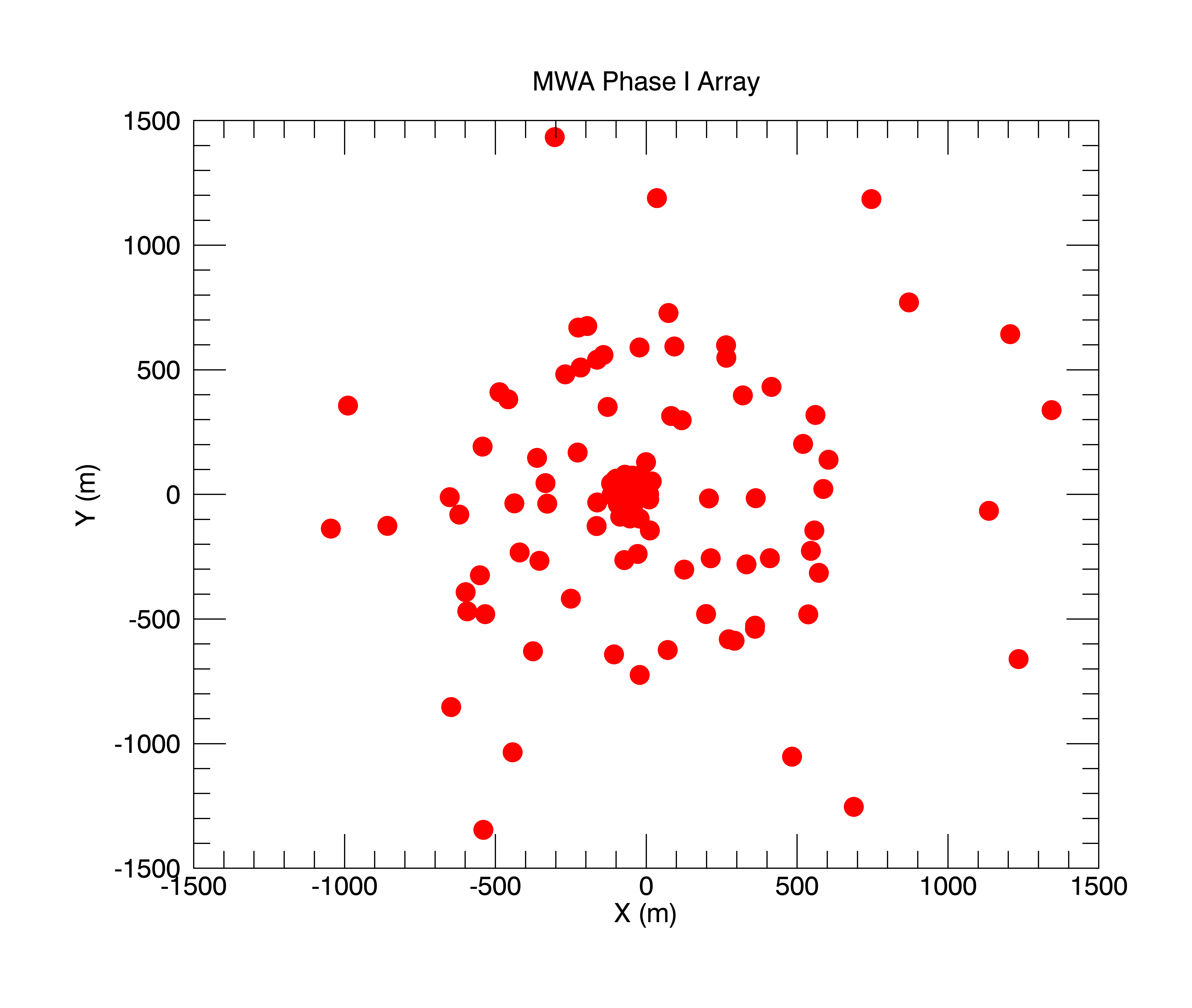}
\includegraphics[width=0.52\textwidth]{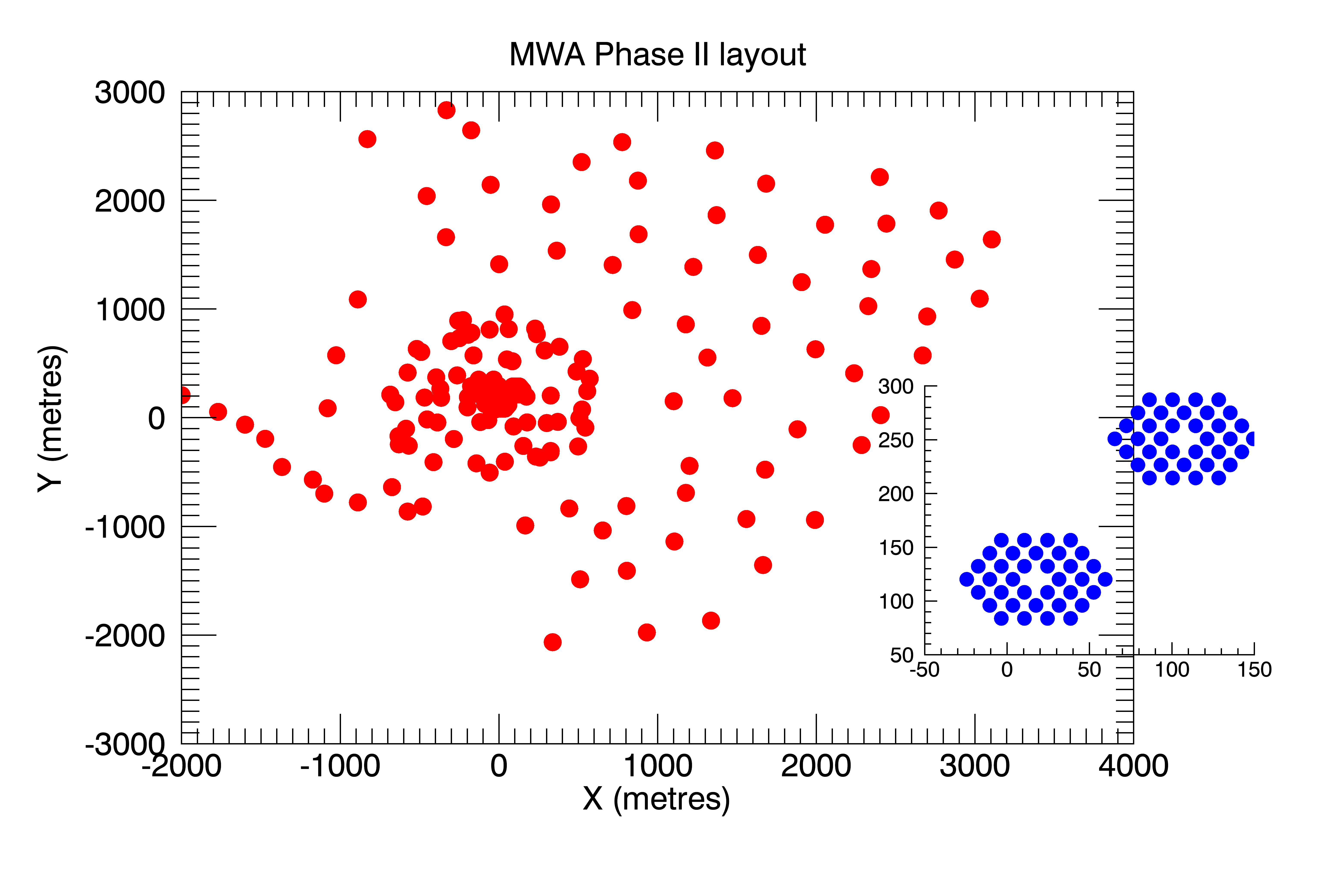}
\caption{Array configurations for the MWA (Phase I, left; Phase II, right, including cutout of hexagonal subarrays (blue)): 128 (256) 4.4~m aperture array tiles spread over 3 (5)~km.}\label{fig:mwa}
\end{figure}

The Murchison Widefield Array (MWA) is a 256-element interferometer in the Western Australian desert. In Phase I of the array, operating from 2013--2016, it was composed of 128 tiles of 16 dual-polarization dipoles, spread over 3~km \cite{tingay13_mwasystem}. Phase II (2016--) expanded the array to 256 tiles, with longer baselines for improved survey science and sky model building (5~km), and two hexagonal sub-arrays of 36 tiles with short spacings available for redundant calibration and improved EoR sensitivity \cite{wayth18}. It operates in two distinct modes: Extended Array (128 tiles with long baselines), and Compact Array (128 tiles with short baselines including two 36-tile redundant subarrays in a hexagonal configuration). The Compact Array is principally used for EoR science (see Figure \ref{fig:mwa}).
The MWA is a general science telescope, with multiple science goals \cite{bowman13_mwascience}. As such, it balances high surface brightness sensitivity on EoR scales, redundant and non-redundant elements, and longer baselines for good imaging capabilities. The instantaneous $uv$-coverage of the MWA is excellent, allowing for science-quality snapshot imaging (2-minute). The MWA is also a wide-field instrument, with a field-of-view of 25~degrees at 150~MHz. This wide field-of-view, combined with the complex frequency-dependent shape of the aperture array primary beam, and analogue electronics, create challenges for data analysis. The two-stage analogue beamformer produces a frequency bandpass that contains 24 coarse channels over a 30.72~MHz bandwidth (chosen from the full bandwidth listed in Table \ref{table:parameters}), with missing regular channels between the coarse bands. This instrumental spectral structure provides a challenge to producing instrumentally-clean output EoR datasets.

Early deployments of the array, with 32 tiles, were used for preliminary science, and to begin to survey the EoR fields \cite{2012ApJ...755...47W}. Upon completion of the 128 tiles, the MWA Commissioning Survey provided the first sky catalogue for use for calibration of EoR data \cite{2014PASA...31...45H}. This work paved the way for the GLEAM survey \cite{wayth15} and catalogue \cite{hurleywalker16}, yielding 300,000 sources in the southern sky. GLEAM provides the basis for the current sky model for point sources in EoR observations, augmented by individual models for extended sources.

In line with developments in concurrent experiments, prior to data acquisition the MWA EoR collaboration focused on relatively simplistic foreground fitting and removal methods, but with an increasing understanding of the signature of smooth-spectrum foregrounds in the wavenumber parameter space of an interferometer \cite{2009ApJ...695..183B,datta10,trott12}.
There are now two primary EoR data calibration and source subtraction pipelines used by the collaboration: the Real-Time System (RTS, \cite{mitchell08}) and Fast Holographic Deconvolution (FHD, \cite{2012ApJ...759...17S}). Both use underlying catalogues of sources that have been generated by cross-matching multiple low-frequency sky catalogues. PUMA \cite{line17} generates an observation-specific sky model of point sources and double sources \cite{procopio15}, and includes shapelet-based and point source-based models for extended sources. The RTS calibrates the data in two steps, both of which rely on a weighted least-squares minimisation: (1) overall direction-independent (flux density and phase) calibration on a full model of 5,000 sources; (2) direction-dependent corrections along the line-of-sight to bright sources. The direction dependent corrections are then applied to sources in the region of the fit, and the 5,000 source sky model is subtracted.
FHD calibration \cite{2012ApJ...759...17S} computationally optimizes the direction-dependent and wide-field imaging steps by pre-computing the mapping from Fourier to real space. FHD relies on an underlying point source sky model \cite{2016MNRAS.461.4151C}, which generates an observation-specific calibration model for $>$10,000 sources based on the GLEAM catalogue and other cross-matched surveys.

Early developments of power spectrum pipelines stemmed from the inverse covariance quadratic estimator framework pioneered in CMB studies \cite{tegmark97}, and applied to theoretical EoR datasets by Liu \& Tegmark in 2011 \cite{liu11}. This work was further developed by Dillon in a series of papers that explored how to bridge some of the differences between the ideal estimator and a physical dataset \cite{dillon15,dillon13}. In particular, Dillon discussed missing data, and large data volumes. An adapted approach was then applied to three hours of MWA data, showing promising results \cite{dillon13}.

One key feature of the optimal quadratic estimator formalism is the whitening of data according to the correlated covariance introduced by the uncertainty on residual foregrounds. This is effectively a down-weighting of data that are heavily affected by foregrounds, thereby improving signal-to-error. Subsequent analysis of a higher redshift dataset was used to estimate the principal eigenmodes of the data in spectral space, identifying these with bright foregrounds \cite{dillon15}.
The covariances of these modes were then used in the estimator to down-weight and decorrelate data, yielding improved limits at $z=6.8$. However, as with commensurate and subsequent work with PAPER that used this technique, it had the large potential to cause bias in the estimates. Re-use of the same dataset to empirically estimate the data covariance, and then fit for it, causes re-substitution bias, a well-known statistical effect where the performance of an estimator can appear much better than it actually is. In this work, Dillon was careful to estimate covariances empirically while omitting the $uv$ cells in question, to avoid bias, however there was still limited information available in the remaining cells. Thus, although this work was careful to not try to subtract the foreground bias directly, use of the empirical covariance in the data weighting, and lack of a full end-to-end simulation to demonstrate no signal loss, makes this approach prone to large bias.  It has not been used to analyze MWA data since the original analysis in \cite{dillon15}.

In a later paper, describing the CHIPS estimator, Trott also developed an inverse covariance quadratic estimator formalism using a model foreground covariance \cite{trottchips2016}. Unlike the empirical approach of earlier work, this does not use the data itself to form the foreground covariance, but a model for the expected spatial and spectral structure of point source foregrounds. However this approach can suffer from similar effects, whereby error in the covariance can propagate into the analysis. Therefore, this inverse foreground covariance has never been applied to data used in publication due to the output's sensitivity to the choice of foreground model.

A second principal power spectrum estimator for MWA EoR analysis, $\epsilon$ppsilon, was independently developed from CHIPS \cite{barry19}. $\epsilon$ppsilon prioritizes the propagation of thermal noise error from the visibilities (with estimates provided by FHD) through to the power spectrum while also providing a suite of diagnostics for assessing the performance of the estimator in a number of domains.

Both $\epsilon$ppsilon and CHIPS (without the foreground covariance weighting) were used in the EoR limit paper led by Beardsley \cite{beardsley16}, which processed 32~hours of MWA Phase I high-band data to power spectrum limits.
At the time, these results were highly-competitive in the field, but the data were clearly still systematic-dominated. At a similar time, Ewall-Wice published the first measurement of upper limit from the Cosmic Dawn (Epoch of X-ray heating, EoX) from 3-hours of MWA data above $z=15$ \cite{ewall-wice16}.

One of the clear outcomes of the early upper limit publications from MWA (and other instruments, particularly LOFAR) was that the data were highly systematic-dominated in modes relevant for EoR, and accumulating more data into the power spectrum estimator would offer no advantage. With this realisation, the MWA collaboration embarked on a two year program to prioritize understanding and treating systematics over processing large datasets, despite more than a thousand hours having been collected by the instrument. This work encompassed (1) improving the sky model (point, extended and multiple sources, \cite{procopio15,trottwayth17_extended}); (2) understanding the impact of calibration choices on residuals and uncertainties \cite{barry16,trottwayth2016,trott17_skala,ewall-wice17,murray17}; (3) improving the primary beam modelling \cite{line18}; (4) developing data quality metrics for data triaging (RFI, ionospheric activity, \cite{jordan16,trott18,wilensky19}), (5) developing and refining redundant and hybrid calibration pipelines for Phase II \cite{li18_redundant,joseph18,byrne19}. A final important step was the development of a full end-to-end simulation to demonstrate that there was no signal loss in the chain from telescope to data product. The results of this work include upcoming EoR limits from re-analysis of Phase I data and new Phase II data, as well as exploration of new techniques for exploring the EoR \cite{trott19_bispectrum}.

A final, key insight from recent work helps to address the current questions in the EoR research field about robustness of any future claimed detection of cosmological signal. Along with confirmation by other telescopes, ability to detect the same signal in independent observing fields, where the foregrounds are different, is crucial. In \cite{trott19_kde}, MWA data from two observing fields was studied with a Kernel Density Estimator to understand the similarities and differences between the statistical structure of data from independent sky areas. This work can lead to a better understanding of robustly discriminating contamination from cosmological signal.

\subsection{Low Frequency Array  - LOFAR}
\begin{figure}[ht]
\centering
\includegraphics[width=0.85\textwidth]{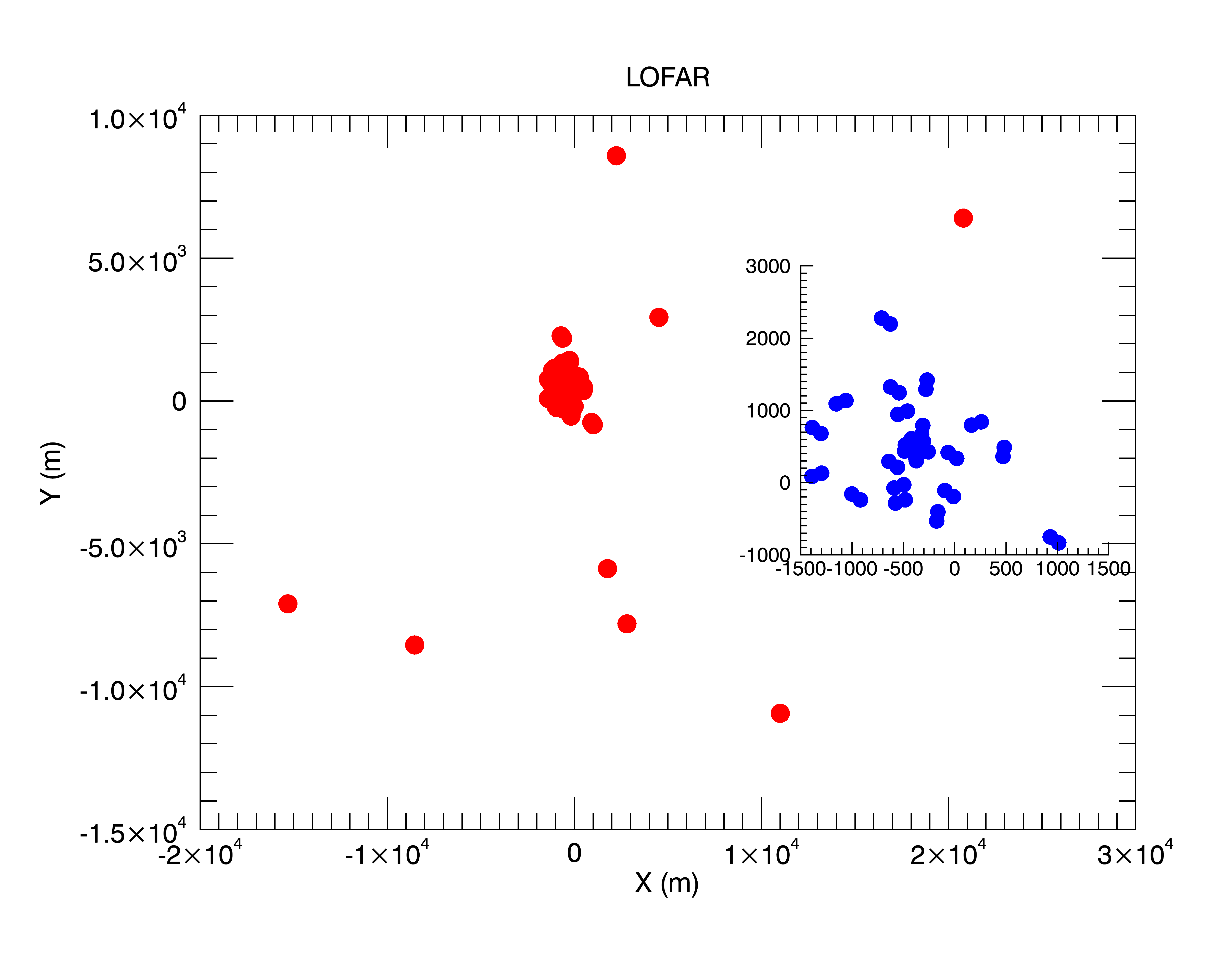}
\caption{Array configuration for the central and inner remote stations of LOFAR (red), and subplot showing the central stations only (blue): 30--40~m aperture array dipole stations spread over tens of kilometres.}\label{fig:lofar}
\end{figure}

LOFAR is a composite aperture array low-frequency radio interferometer. It has two primary station types; the High-Band Antennas (HBA, 120--190~MHz) and Low-Band Antennas (LBA, 30--90~MHz). Both station types have been used for EoR and Cosmic Dawn science. Athough LOFAR formally contains baselines of thousands of kilometres to the international stations, it is only the Dutch-based stations that are used for actual EoR measurements. Figure \ref{fig:lofar} shows the central stations (blue cut out) and the nearest remote stations (red).

The LOFAR latitude allows for circumpolar observations with long winter nights. As such, one of the primary observing fields is the North Celestial Pole, which can be observed for more than 12 hours in the winter months.
Early work with the LOFAR EoR Key Science Project focussed on foreground mitigation, choice of observing fields, and data analysis methodology. As with many of the published papers in this early epoch, foregrounds in \cite{2008MNRAS.389.1319J} were modelled to be subtracted with a simple smooth fitting function. This work is notable because it provided realistic models for a range of different foreground components, and included discussion of the treatment of polarized foregrounds.

\cite{2010MNRAS.409.1647J} extended the work from 2008 to focus on simulations of Faraday Rotation from polarized Galactic foregrounds. FR rotates the phase of the intrinsically-smooth foreground component yielding spectral structure that may mimic the EoR signal. In this work, Jeli{\'c} shows the effect of inaccurate data calibration on polarized emission, which can imprint total intensity structure if the polarized instrumental response is incorrect.


The SAGE algorithm (Space Alternating Generalized Expectation Maximization) was first introduced in 2011 by \cite{2011MNRAS.414.1656K}, and provides the basis for all calibration of LOFAR EoR datasets to the present day. Based on the well-known Expectation Maximization (EM) algorithm, which iteratively fits for calibration parameters (maximizes the likelihood with respect to a set of parameters and then alters the parameters to find a new likelihood) when the underlying system model contains unobserved variables, SAGE extends the traditional least-squares fitting to allow for more model flexibility, and improved convergence and efficiency. Part of the SAGE algorithm performs direction-dependent calibration towards clusters of sources on the sky, thereby allowing for ionospheric distortion of the sky model.

Detailed total intensity and polarized imaging of the LOFAR EoR fields were presented in \cite{2013A&A...550A.136Y} and \cite{2014A&A...568A.101J}. The North Celestial Pole field allows for deep, long winter nighttime observations, and was shown to be able to be calibrated over long integrations. Observations of low Faraday depth structures in the ELAIS-N1 field yielded structures that would be problematic for EoR science if the degree of polarization leakage into total intensity exceeded 1\%. This quantification of the accuracy required of instrumental polarization models was the first of a set of papers that explored polarized signal and leakage for EoR science. Thus far, the LOFAR EoR collaboration has undertaken the most extensive work to quantify the impact of polarization leakage, while the MWA collaboration has made some observations of polarized emission in their data (\cite{2016ApJ...830...38L}, \cite{2017PASA...34...40L}, \cite{2013ApJ...771..105B}). In \cite{2015MNRAS.451.3709A} and \cite{2016MNRAS.462.4482A}, Asad and colleagues first studied the polarized emission in the 3C196 EoR field, finding them to be localized around a small Faraday depth, and quantified the leakage, and then studied the accuracy of the LOFAR polarized beam model to be able to limit leakage into total intensity. Given the level of polarized to total intensity, and a beam model accurate to 10\% at the field centre, the leakage and subsequent spectral structure would be acceptable for EoR science. Finally, \cite{2018MNRAS.476.3051A} considered the more problematic impact of wide-field polarization leakage on the EoR power spectrum. Far from the field centre, the primary beam models are less accurate, and sources imprint additional spectral structure due to the chromaticity of an interferometer. In these cases, bias was found to persist in the EoR power spectrum.

In early work on fitting foregrounds, Harker and colleagues \cite{2010MNRAS.405.2492H} discussed the use of Wp smoothing as a non-parametric method for fitting a smooth function, based on limiting the number of inflection points in the fit. Further, they discuss the systematic errors introduced by the fitting routine, methods for estimating these, and for accounting for them in the final uncertainties. This approach is used again in the work of \cite{2018MNRAS.478.3640M} for the Gaussian Process Regression fitting, and also is used generically in the PAPER analysis to try to understand signal loss.
There, and elsewhere, use of the same dataset to empirically estimate the bias, and then to correct it, leads to resubstitution bias, underestimate of bias and signal loss.

In a series of papers, Chapman and collaborators explored novel approaches to fitting and removing foreground signal from image-based datacubes. In \cite{2012MNRAS.423.2518C}, they introduced the FastICA technique, as a non-parametric method that estimates independent foreground components and their mixing for each image pixel and frequency. The advantage of such methods is that they do not rely on any a priori knowledge of the signal, but instead only assume that the full signal can be represented by a small number of components (sparsity), thereby allowing for good estimation with a given dataset. The disadvantage lies in the sensitivity of results to the number of components the user chooses that the data should contain, and the potential for signal loss if the projection of the estimated components onto the EoR signal is non-negligible. ICA generically ignores stochastic components, thereby isolating smooth components of the data, and minimizing Gaussianity, thereby enforcing smoothness in the fitted components. Beyond ICA, a generalized method (GMCA; Generalized Morphological Component Analysis) was applied (\cite{2013MNRAS.429..165C,chapman14}) to use an underlying blind wavelet decomposition of the components, combined with the sparsity and mixing model methodology of the ICA method. As with other methods when the underlying structure, spatial distribution and amplitude of the cosmological signal and foreground components is unknown, the potential for signal loss is present. The ICA and GMCA methods both assume that the cosmological signal has negligible amplitude and is absorbed in the noise. Structural deviations from this assumption can lead to signal loss.

In a new approach to foreground treatment, Mertens and colleagues discuss use of the well-known Gaussian Process Regression (GPR) technique to fit for foregrounds using only an understanding for the spectral data covariance of different components \cite{2018MNRAS.478.3640M}. Unlike parametric methods that assume an underlying model, GPR (an extended version of kriging, which interpolates data based on their known covariance properties) relies only a statistical separation of foreground and cosmological signal via their spectral correlation lengths. This method also suffers from the potential for cosmological signal loss, but the authors attempt capture the potential bias statistically through increased noise. The ultimate utility of this approach has yet to be demonstrated on a large dataset at the time of the writing.

Along with the power spectrum as a measure of the signal variance as a function of spatial scale, the variance statistic was explored with simulations in \cite{2014MNRAS.443.1113P}. The variance of the brightness temperature, as the wavenumber integral over the power spectrum, quantifies the variability in the cosmological signal on the imaging scale (autocorrelation function). Although it provides limited cosmological information, detection of this variance can be theoretically obtained with fewer observing hours. Bayesian power spectrum extraction techniques were also explored in \cite{2015MNRAS.452.1587G}, with a view to allowing for a spatially-smooth component to capture the unmodelled diffuse emission in the NCP field. Like other instruments, the data calibration and foreground models were limited to point and extended sources, with the complex diffuse emission difficult to measure and model. Increasingly, the impact of this incomplete sky model has become apparent.

In a landmark paper published by Patil and colleagues in 2016 \cite{patil16} the source of `excess noise' and diffuse emission suppression in LOFAR data were studied. Excess noise is the identification of increased noise levels in the data post-calibration compared with expectations of thermal noise and Stokes V measurements. Ultimately, the lack of a diffuse model in the calibration sky model allowed for this signal to be absorbed into the gain calibration solutions, thereby yielding a direction-dependent bias and noise in the residual data. To address this problem, the short baselines containing the majority of the diffuse emission could be excluded, however this leads to increased noise on these scales (due to statistical leverage; effectively this amounts to additional flexibility in the gain solutions on these scales because they are not used in the modelling). This work was undertaken contemporaneously with that of \cite{barry16} and \cite{trottwayth2016}, which both studied the effect of incomplete sky models and spectrally varying bandpass parameters on calibration and residual signal. The combined outcome of these studies is an understanding of the impact of sky model incompleteness, the need to enforce spectral correlation (e.g., regularization as in SAGECal or smooth model fitting) for calibration parameter fitting, and the approaches to calibration that can mitigate these.

Further exploration of the impact of calibration frameworks and data treatment were then explored in \cite{2019MNRAS.483.5480M} and \cite{2019MNRAS.484.2866O}, with a view to having a complete understanding of the end-to-end data processing of LOFAR EoR data on the path to a detection. Unlike in the previous ten years before real observations were undertaken and thermal noise sensitivity was seen to be the major impediment for EoR detection, the field has come to appreciate the crucial roles of unbiased calibration, sky model completeness, and foreground treatment without cosmological signal loss.

The culmination of the lessons learned from statistical leverage and incomplete sky models was applied to two fluctuation upper limit papers published since 2017. In \cite{2017ApJ...838...65P}, Patil and colleagues presented competitive results from a small set of data ($\sim$10 hours) at $z = [9.6 - 10.6]$, with the best limit of (59.6 mK)$^2$. This work reported an excess variance, in line with previous discussions, and the use of Stokes V power to remove noise power. At higher redshifts (lower frequencies), Gehlot and colleagues \cite{2018arXiv180906661G} reported upper limits above $z=20$, with use of the Gaussian Process Regression foreground fitting technique introduced by \cite{2018MNRAS.478.3640M} for EoR science.

Beyond the power spectrum, LOFAR has explored other tracers of the neutral hydrogen temperature field, namely the ability to produce low angular resolution images \cite{2012MNRAS.425.2964Z} and the 21~cm Forest \cite{2013MNRAS.428.1755C}. LOFAR like other current instruments, does not have the sensitivity to directly image neutral hydrogen bubbles at the instrumental resolution. However, by lowering the resolution of images (thereby improving the radiometric noise), Zaroubi and colleagues argue that the largest of bubbles may be detectable at low signal-to-noise ratio on the largest of scales late in reionization. The ability to detect the 21~cm Forest (absorption of continuum radio emission along the line-of-sight to high redshift AGN due to intervening neutral gas) remains a challenge and aim of many current interferometers. Ciardi and colleagues showed that LOFAR would have the ability to detect an absorption feature under ideal conditions. Unlike the statistical detection of the power spectrum of temperature fluctuations, the absorption signal amplitude is determined by the astrophysical conditions close to the gas, namely gas kinetic temperature. Cold gas is able to absorb light more readily than heated gas. The failure of this method to date is primarily due to the lack of any known high-redshift radio-loud AGN ($z>6$). Given the sensitivity of current instruments, a source with flux density exceeding 10~mJy and cold neutral gas would be required. It is likely that the arrival of SKA will provide both the sensitivity and the detection (and confirmation) of high-redshift radio-loud AGN to be able to undertake this experiment. Of the current interferometric experiments, only LOFAR has sufficient sensitivity to be able to attempt this experiment at all.

The utility of extracting higher-order statistics of the 21~cm brightness temperature field were explored in a simulation study of foreground-subtracted image cubes by \cite{2009MNRAS.393.1449H}. Again, the ability to smoothly treat and remove foregrounds placed the burden of detection on pure noise considerations.

\subsection{Precision Array for Probing the Epoch of Reionization - PAPER}
The PAPER experiment was designed as a testbed for developing novel 21~cm cosmology analysis techniques.  PAPER antennas were chosen to be small, single dipoles on elevated ground-screens to enable reconfiguration of the array, and the system used a flexible digital correlator architecture that could scale as the number of antennas grew \cite{parsons08}.  The small antenna sizes was also chosen to limit the frequency evolution of the antenna response over the instrument's $110-180$\, MHz of usable instantaneous bandwidth.  The design and results from an initial 8-station deployment of PAPER in Green Bank, WV, USA were described in \cite{parsons10}.

In its earlier stages, PAPER deployed its antennas in configurations designed for imaging, including a single-polarization 16-element 300~m diameter ring in Green Bank used for primary beam measurements in \cite{pober12}.  While the Green Bank array was upgraded to a single-polarization, 32-element array, all subsequent publications came using arrays deployed at the SKA-SA site in the Karoo, South Africa.  Highlights of early PAPER studies include the creation of a 145\, MHz Southern hemisphere sky-catalog using a single-polarization, 32-element array \cite{jacobs10} and a study of the radio galaxy Centaurus A using a single-polarization, 64-element array \cite{stefan10}.  In both of these cases, the elements were deployed in a randomized configuration over a circle of 300~m diameter to maximize $uv$ coverage.

In 2012, however, members of the PAPER team developed what is now referred to as the ``delay spectrum" approach for measuring the 21~cm power spectrum.  In the delay spectrum approach, visibility spectra from individual baselines are Fourier transformed and cross-multiplied. \cite{parsons12a} demonstrated how these delay spectra can be used as estimates of the 21~cm power spectrum, without ever combining visibilities and making an image.  \cite{parsons12a} also provided sensitivity estimates for the delay spectrum approach using a 128-element PAPER array.  \cite{parsons12b} then demonstrated how 21~cm foregrounds isolate into what is now commonly referred to as ``the wedge" and included the effects of foreground contamination in the sensitivity study.  One consequence of the delay spectrum approach is a higher noise level than alternative approaches: power spectra estimated from individual baselines are averaged together, as opposed to coherently combining all the visibilities and forming a single power spectrum, so noise fluctuations average down more slowly.  To make-up for this sensitivity sacrifice, \cite{parsons12a} proposed using a ``maximum redundancy" configuration, in which antennas are arranged to create multiple copies of the same baseline spacing.  These redundant baselines can then be averaged together before squaring, helping the noise level to integrate down faster.  Although redundant layouts drastically reduce imaging fidelity, the delay spectrum approach does not requiring imaging and so is, in principle, not affected by this consequence.

The decision was made to reconfigure the PAPER array and test the delay spectrum technique in a maximum redundancy layout. However, a short data set in a single-polarization, 64-element ``minimum redundancy" (i.e. random layout) with a 300~m diameter was collected and used to make delay spectra from a range of baseline lengths and orientations in \cite{pober13}.  This analysis demonstrated good isolation of foreground emission to the wedge in 2D cosmological $k$-space, suggesting the promise of the delay spectrum technique.

\cite{parsons14} presented the first deep power spectrum limits from a dual-polarization, 32-element, maximum redundancy array (a grid of 8 columns and 4 rows, with a column spacing of 30 meters and a row spacing of 4 meters).  Just over 1000 hours of data were used in the analysis.  In addition to the basic delay spectrum formalism, \cite{parsons14} introduced two additional analysis techniques: redundant calibration \cite{wieringa92,liu10}, which was enabled by the redundant layout of the array, and a new technique for removing off-diagonal covariances between redundant baselines.  These same techniques were applied to the same data over a range of redshifts in \cite{jacobs15}.

\begin{figure}[ht]
\centering
\includegraphics[width=0.85\textwidth]{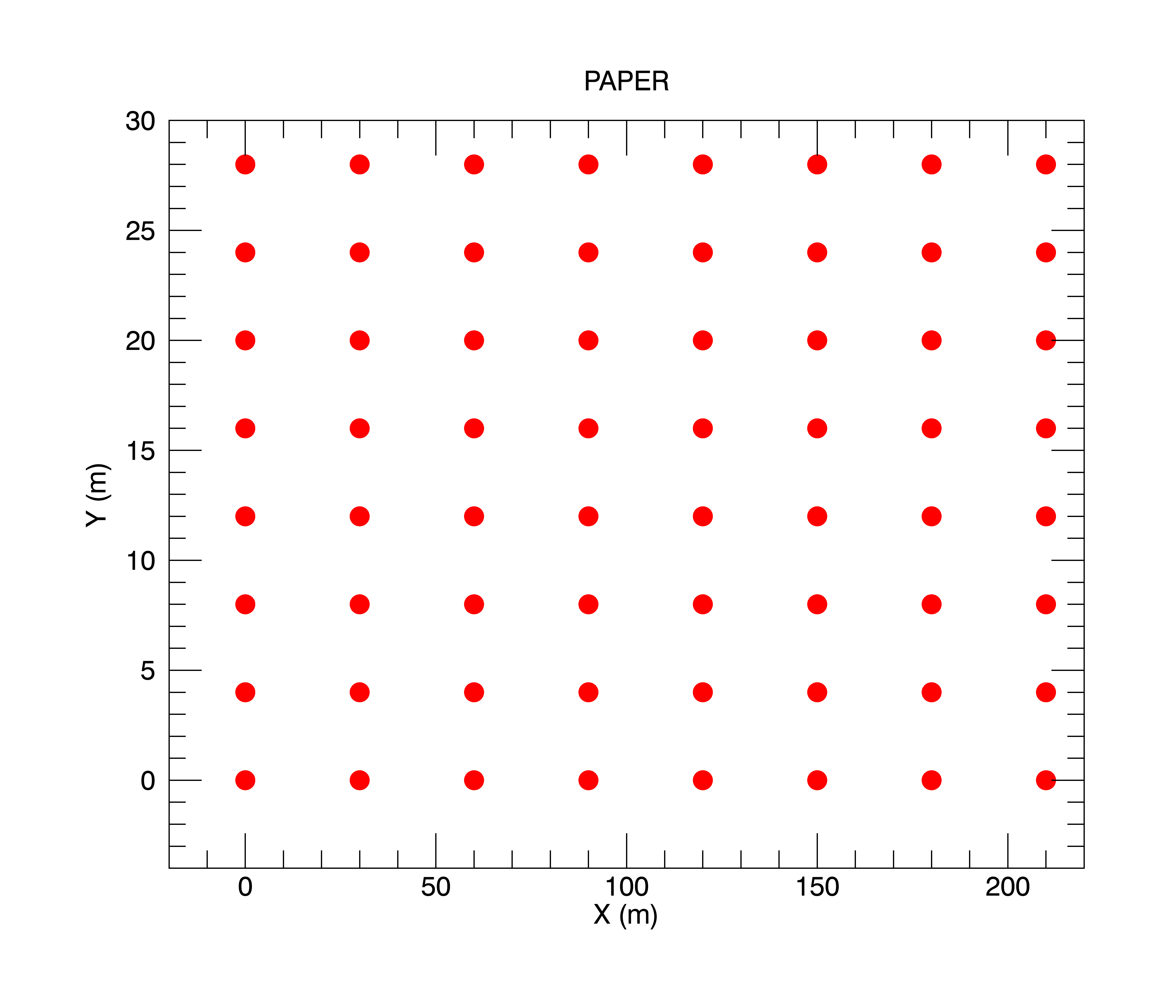}
\caption{Array configuration for the PAPER-64 maximally-redundant array: 64 single dipoles spread over 210~m.  Note the distinctly different scales on the $x$ and $y$ axes.}\label{fig:paper}
\end{figure}

The techniques of \cite{parsons14} were then applied to a new, $1000+$ hour, dual-polarization, 64-element PAPER data set in \cite{ali15} (the layout of which is shown in Figure \ref{fig:paper}).  This analysis improved upon the redundant calibration technique by using the OMNICAL package \cite{zheng14}, replaced the off-diagonal covariance removal technique with an inverse covariance weighting approach using empirically estimated covariance matrices (similar to \cite{dillon15}) and applied a new technique known as ``fringe rate filtering" (described in \cite{parsons16}).  At the time, the \cite{ali15} limits on the 21~cm power spectrum were believed to be the most stringent to be published and were followed by two separate publications using their measurement to constrain the temperature of the IGM at $z=8.4$ \cite{pober15,greig16}.

However, re-analysis of the \cite{ali15} data by \cite{cheng18} revealed a critical error in the analysis: empirically estimated covariance matrices are correlated with the data, and weighting by them can bias the recovered signal low.  (As described in \S\ref{sec:gmrt}, this bias has frequently been referred to as ``signal loss" --- the idea that an analysis technique can remove 21~cm signal along with foregrounds.)  In practice, the signal loss in the PAPER analysis was \emph{very} large (nearly four orders of magnitude of potential EoR signal was suppressed) due to the fringe-rate filtering technique that reduced the number of independent samples used to estimate the covariance matrix.  Although the analysis in \cite{ali15} attempted to estimate signal loss using injection of mock EoR signals into the data, their method missed potential loss caused by data-signal cross terms in the covariance matrix and thus concluded that the original analysis was effectively lossless.  Incorrect estimates of both the theoretical and observed noise levels in the data also contributed to the belief that the analysis of \cite{ali15} was sound.

In light of the analysis in \cite{cheng18}, all of the PAPER results in \cite{parsons14,jacobs15,ali15} are considered to be invalid and do not place meaningful limits on the 21~cm signal.\footnote{Although \cite{parsons14} did not use the inverse covariance weighting that was the main source of the problem in \cite{ali15}, its covariance removal technique has not been robustly vetted for signal loss and thus the results are considered suspect at best.} A re-analysis of the full \cite{ali15} data set using a lossless analysis is forthcoming, but the limits are not expected to be near the same level as \cite{ali15}.  The PAPER experiment also collected two years of data with a dual-polarization, 128-element array, but due to an increased amount of instrument systematics and failures in the aging system, these data are not expected to be published.

The delay spectrum approach does not allow for high accuracy polarization calibration, which needs to be performed in the image domain.  Theoretical studies of the effect of Faraday rotated (i.e. frequency-dependent) polarized emission on the delay spectrum technique were presented in \cite{moore13} and \cite{nunhokee17} and studies using PAPER data were performed in \cite{moore17} and \cite{kohn16}.  Overall, the effect of polarized emission on the delay spectrum approach can be quite significant, but the overall amplitude is uncertain as there are few constraints on the polarization properties of the 150~MHz sky at the angular scales probed by PAPER.  Ionospheric Faraday rotation can also attenuate the polarized signal in data sets averaged over many nights \cite{martinot18}.

\section{Published results}
\label{sec:results}
Here we collate the published best limits at each redshift from the current experiments (Table \ref{table:limits}). PAPER measurements have been omitted. Despite the current published values, there are publications in peer-review now for LOFAR, PAPER, and MWA improving on these results.
\begin{table}[ht]
\centering
\begin{tabular}{|c|c|c|c|c|}
\hline
Facility & $z$ & $k$ ($h$Mpc$^{-1}$) & Upper limit (mK)$^2$ & Ref. \\
\hline
MWA & 12.2 & 0.18 & 2.5$\times$10$^7$ & \cite{ewall-wice16}\\
MWA & 15.35 & 0.21 & 8.3$\times$10$^8$ & \cite{ewall-wice16}\\
MWA & 17.05 & 0.22 & 2.7$\times$10$^8$ & \cite{ewall-wice16}\\
MWA & 7.1 & 0.23 & 2.7$\times$10$^4$ & \cite{beardsley16}\\
MWA & 6.8 & 0.24 & 3.0$\times$10$^4$ & \cite{beardsley16}\\
MWA & 6.5 & 0.24 & 3.2$\times$10$^4$ & \cite{beardsley16}\\
MWA & 9.5 & 0.05 & 6.8$\times$10$^4$ & \cite{dillon15}\\
LOFAR & 10 & 0.053 & 6.3$\times$10$^3$ & \cite{patil16}\\
LOFAR & 9 & 0.053 & 7.5$\times$10$^3$ & \cite{patil16}\\
LOFAR & 8 & 0.053 & 1.7$\times$10$^4$ & \cite{patil16}\\
LOFAR & 23 & 0.038 & 2.1$\times$10$^8$ & \cite{2018arXiv180906661G}\\
GMRT & 8.6 & 0.5 & 6.2$\times$10$^4$ & \cite{paciga13}\\
\hline
\label{table:limits}
\end{tabular}
\caption{Best two sigma upper limits on the EoR and Cosmic Dawn power spectrum for each experiment. Only the lowest limits have been reproduced.}
\end{table}

\section{Current challenges}
\label{sec:challenges}
21~cm experiments consist of many components, from the analog telescope design through to power spectrum estimation algorithms.  One clear lesson from first generation experiments is that no one aspect of the system can provide the necessary 1-part-in-$10^5$ dynamic range required to detect the 21~cm signal; rather, the burden needs to be spread across the components of the experiment, alleviating the demands on each of the other components.  In this section, we briefly review what we consider five key areas where 21~cm experiments continue to innovate: (1) analog instrument design; (2) data quality control; (3) calibration; (4) foreground mitigation and the associated potential for signal loss; and (5) end-to-end validation of analysis pipelines.

\textbf{1. Analog instrument design}.  One of the major challenges for 21~cm cosmology experiments is to remove any spectral structure introduced by the instrument that might otherwise mix smooth spectrum foregrounds into the spectral modes occupied by the cosmological signal.  One seemingly straightforward approach is to limit the amount of spectral structure in the instrument response through careful analog design.  Initial specifications on the HERA system design were to limit spectral structure to a level that would enable the delay spectrum technique without any additional calibration or analysis requirements; however, further study has shown that the HERA design does not meet this stringent specification and will need data analysis algorithms to also model and remove spectral structure from the instrument \cite{deboer16}.  The push to larger bandwidths (e.g. $50-250$\, MHz for HERA and $50-350$\, MHz for the SKA) adds to the challenge of constructing a single instrument with a smooth spectral response over a large range of wavelengths.  Analysis with the MWA has also demonstrated how reflections in the analog system can contaminate modes of the EoR power spectrum, suggesting that more stringent specifications on impedance matches and cable lengths are necessary for future instruments \cite{barry16,ewall-wice17}.

\textbf{2. Data Quality Control.} Given the extreme brightness of human-generated radio signals compared to the 21~cm signal, only a very small number of contaminated measurements are enough to significantly affect the analysis of a large data set.  The ``gold standard" for identifying radio frequency interference, AOFlagger \cite{offringa12}, is used by both LOFAR and the MWA.  However, additional quality metrics can still catch corrupted data that slips by this first round of flagging, including ultra-faint, broad-band digital TV transmission \cite{wilensky19} and effects due to ionospheric weather \cite{trott18,jordan16}.  As interferometers grow in size, the large data rates may also require computationally faster algorithms for data quality checks \cite{kerrigan19}. \cite{offringa19} also demonstrate how even flagged RFI can affect power spectrum analysis if care is not taken.

\textbf{3. Calibration.} Instrument calibration is often regarded as the greatest challenge for existing and future 21~cm experiments.  While both the analog design and the methodology used for power spectrum estimation can ease calibration requirements \cite{morales19}, experiments still need to control the spectral response of their telescopes over wide bandwidths at a level unprecedented in radio astronomy.  Typically, antenna-based gain calibration is performed by forward-modeling visibilities and minimizing the difference with the observed data; however, \cite{barry16}, \cite{patil16}, and \cite{trottwayth2016} demonstrate that without additional constraints, calibration performed with an incomplete sky-model can lead to spurious spectral structure in the calibration solutions that can both overwhelm or remove the EoR signal.  Redundancy based calibration has been viewed as a promising alternative because it does not reference a sky-model; however, recent work has shown that a sky model is still required to constrain the degeneracies inherent in redundant calibration, and that the same kind of contamination can affect the power spectrum as in sky-based calibration \cite{byrne19,li18_redundant,joseph18}. Calibration of the primary beam response of the instruments is also a major challenge, and several options have been explored, including sky-based calibration \cite{pober12}, using satellite broadcasts \cite{neben15,neben16,line18}, and with drones flying transmitters \cite{jacobs17}.

\textbf{4. Foreground mitigation.} Fundamentally, the real challenges at the heart of 21~cm cosmology come from the intrinsic brightness of the foreground emission.  While much of the work to date focuses on removing the instrument response from the foreground spectra, most current experiments use some form of foreground mitigation to help isolate or remove foregrounds.  Many distinct approaches have been developed, which can be broadly classified as either ``foreground avoidance" and ``foreground subtraction."  Foreground avoidance methods attempt to isolate foregrounds into the wedge and minimize bleed into the EoR window; power spectra are then only estimated from within the EoR window.  Examples of avoidance techniques includes the wide-band iterative deconvolution filter used in PAPER analyses \cite{kerrigan18} and the inverse covariance weighting techniques also used by PAPER \cite{cheng18}.  Foreground subtraction, on the other hand, attempts to remove specific models of the foregrounds --- using either real sky catalogs or parametric models for their spectra --- while leaving the 21~cm unaffected.  Examples of foreground subtraction including the point-source forward modeling and subtraction performed by FHD \cite{barry19} and the spectral based fitting methods used by LOFAR \cite{chapman14,2018MNRAS.478.3640M}.  It is worth stressing that while these techniques have historically been developed in the context of specific experiments, they are more generally applicable; see \cite{kerrigan18}
for an example of PAPER-developed techniques applied to MWA data and MWA-developed techniques applied to PAPER data.

One of the greatest risks of foreground removal is the inadvertent removal of 21~cm signal, i.e., signal loss.  Although many techniques have been developed using frameworks where signal loss is not expected, due to a presumed orthogonality of the foreground description and 21~cm signal basis, there are subtle challenges that arise when faced with a need to achieve five orders of magnitude of dynamic range.  While cross-terms between the foreground and signal might have an expectation value of 0, there are still only a finite number of samples going into the analysis, and these cross terms will not have converged to their expectation value --- as was the case in the PAPER analysis of \cite{ali15}.

\textbf{5. Validation.} One of the last major challenges for current and future experiments is to rigorously test foreground removal and other analysis algorithms --- ideally as part of complete pipeline and not as an independent step --- to confirm that 21~cm signal is not being biased or removed.  And although foreground removal seems like the step most likely to cause signal loss, it is certainly not the only place that needs further scrutiny.  In light of the PAPER retractions, 21~cm experiments are realizing the importance of simulation-based analysis vetting --- ideally with independent, third-party simulations.  Many inteferometric simulators exist, including CASA, PRISim \cite{thyagarajan15a}, OSKAR, and pyuvsim \cite{lanman19}.  In turn, it has become important to test the simulators against each other, to verify that they achieve the requisite precision for 21~cm cosmology.  These validation efforts can be slow and painstaking, but as experiments push closer to a first detection of the 21~cm signal, they have become more vital than ever. The other avenue for verification is with other instruments and other pipelines providing independent analysis. Use of multiple observing fields can also show robustness to foreground treatment \cite{trott19_kde}.

\section{Prospects for the future}
\label{sec:prospects}

\subsection{Current instruments}
MWA and LOFAR are both currently pursuing deeper limits. Armed with new calibration and analysis, and critically, a deeper understanding of the effects of different processing approaches, the level of systematics in data are reduced, and more data can be processed to reduce noise. At this stage, it is difficult to predict whether systematics will remain at deeper levels, and if the fundamental limitations of the instrument will preclude a detection. While the reported detection of the Cosmic Dawn global signal from the EDGES experiment \cite{bowman18} suggests that the spatial power spectrum amplitude may be larger than expected, this is highly uncertain, and the flexibility in possible strengths of the signal in the EoR emission part of the spectrum could help or hinder a detection by LOFAR and MWA. Pursuit of the Cosmic Dawn signal from 75--100~MHz observations with the MWA and LOFAR is also underway, but that introduces even greater challenges of large fields-of-view and poor extended source models at those frequencies.

\subsection{Future instruments}
The SKA and HERA offer the future vision for EoR and Cosmic Dawn science. Like LOFAR and MWA, SKA is a general science instrument, being able to produce its own sky model and calibration framework, while needing to balance design with the other science aims of the observatory. HERA, like PAPER, is a custom EoR instrument, being able to design with a complete focus on EoR science, likely requiring external information to provide a full end-to-end calibration and source subtraction element.

The low-frequency telescope of the SKA Observatory, SKA-Low, will be centred at the Murchison Radioastronomy Observatory in Western Australia, on the same radio quiet site as MWA, ASKAP, EDGES and BiGHORNS \cite{koopmans15,dewdney16}. Despite being designed for 512 38~m stations (256 dual-polarization dipoles in each station) spread over $>$40~km, the core region will contain $>$200 stations within the central 1~km, with exceptional surface brightness sensitivity for EoR and CD science. With a frequency range available down to 50~MHz, the CD will be accessible to $z=27$, with sub-stations able to be formed to produce the wider fields-of-view and shorter baselines required for early times. With its exceptional imaging capabilities, SKA-Low aims to pursue power spectrum, direct imaging (tomography) and 21~cm Forest studies. The prices to be paid for this highly-capable instrument are the complexity of the data and instrument, and the large data volumes that will be produced from the telescope, and is therefore faces a more severe version of the challenges currently experienced by multi-purpose dipole arrays such as MWA and LOFAR.

HERA \cite{deboer16} is a smaller instrument (although still significantly larger than any of the existing instruments) being constructed in South Africa. It comprises 350 14~m dipole elements spread over $<$1~km (331 in a 320~m core) for high EoR sensitivity and moderate imaging and calibration needs (19 outriggers). It will primarily pursue the statistical exploration of the EoR and CD using the delay spectrum technique, with some hope for imaging capability and alternate power spectrum analyses.

\subsection{Future analyses}
Although the spatial power spectrum is the primary data product of most current EoR 21~cm experiments, there are other avenues of pursuit to explore this first billion years of the Universe, including an integrated product (the variance statistic, \cite{patil14}). Direct imaging is beyond the capability of current instruments, demanding a high surface brightness sensitivity and thousands of hours. This will be pursued by the future SKA~\cite{koopmans15}. However, there are other statistics that can be pursued through the 21~cm line, and also the opportunity for cross-correlating the signal with other tracers of early Universe evolution. The benefit of the latter approach is that the systematic errors may be different between the two tracers, offering an advantage over 21~cm alone.

At early times, the brightness temperature of the 21~cm line, relative to the CMB traces the matter power spectrum, and is highly Gaussian, but at later times the evolution of ionised bubbles dominates the spatial fluctuations and the signal is expected to have non-zero higher order terms \cite{furlanetto06,mcquinn06,eisenstein99}. The shape of the temperature distribution function evolves with time and spatial scale, and differs for different underlying models of the evolution of the Universe. As such, probing these non-Gaussian components can provide complementary information to the power spectrum, which, by design, only captures information in the second moment of the distribution~\cite{wyithe07}.

The bispectrum measures the three-point correlation function, and has been shown to encode non-Gaussianity. In early work to study the expected sensitivity of 21~cm experiments to the bispectrum,~\cite{yoshiura15} computed theoretical expectations for a range of instruments, under the assumption of thermal noise only. More sophisticated recent work included the effects of calibration and foregrounds on the ability to detect the signal. In~\cite{trott19_bispectrum}, two bispectrum estimators were developed to take a practical approach to estimation with real data, and were applied to 20 hours of data from Phase II of the MWA.\@ This work discussed some of the advantages and challenges of doing such an experiment with real data.

Cross-correlation studies from the early Universe offer the potential for new astrophysical insight and reduced observational biases and errors. In the context of the MWA,~\cite{yoshiura19} used data to explore the cross-correlation of the 21~cm image from the EoR-0 observing field, and the CMB field measured by Planck. An additional tracer that can be used is the population of high-redshift LAEs, which are observable in ionised regions~\cite{yoshiura18,kubota18,hutter17}. The SKA's Synergy group is exploring the potential for multi-facility observations, including the exciting prospects available with WFIRST \cite{hutter19}.


\bibliographystyle{plain}
\bibliography{ms_trottnew}


\end{document}